\documentclass[aps,prd,floats,floatfix,superscriptaddress,twocolumn,PRD,nofootinbib,amsmath,amssymb,amsfonts,longbibliography,preprintnumbers]{revtex4-2}

\usepackage[T1]{fontenc}
\usepackage[utf8]{inputenc}

\usepackage[dvipsnames, usenames]{xcolor}
\usepackage{graphicx}
\usepackage[normalem]{ulem}
\usepackage{bm}
\usepackage{appendix}
\usepackage{microtype}
\usepackage[english]{babel}

\definecolor{linkcolor}{rgb}{0.0,0.3,0.5}
\usepackage[hypertexnames=false, unicode, colorlinks=true, linkcolor=linkcolor, citecolor=linkcolor, filecolor=linkcolor,urlcolor=linkcolor]{hyperref}

\usepackage{array}
\newcommand{\PreserveBackslash}[1]{\let\temp=\\#1\let\\=\temp}
\newcolumntype{C}[1]{>{\PreserveBackslash\centering}p{#1}}
\newcolumntype{R}[1]{>{\PreserveBackslash\raggedleft}p{#1}}
\newcolumntype{L}[1]{>{\PreserveBackslash\raggedright}p{#1}}

\renewcommand{\vec}[1] {\bm{#1}}
\newcommand{\vhat}[1]{\vec{\hat{#1}}}
\newcommand{\kpivot}{k_0}

\definecolor{darkred}{RGB}{175,0,0}
\definecolor{darkblue}{RGB}{14,0,185}
\definecolor{lightblue}{RGB}{100,149,237}

\begin{document}
\preprint{UTWI-26-2024}

\title{Searching for blue in the dark}

\newcommand\alvisehome{
\affiliation{Dipartimento di Fisica Galileo Galilei, Universit\` a di Padova, I-35131 Padova, Italy}
\affiliation{Istituto Nazionale di Fisica Nucleare Sezione di Padova, I-35131 Padova, Italy}
\affiliation{Istituto Nazionale di Astrofisica Osservatorio Astronomico di Padova, Italy}
}

\newcommand\jessiehome{
\affiliation{Dipartimento di Fisica Galileo Galilei, Universit\` a di Padova, I-35131 Padova, Italy}
\affiliation{Istituto Nazionale di Fisica Nucleare Sezione di Padova, I-35131 Padova, Italy}
}

\newcommand\elehome{
\affiliation{Dipartimento di Fisica Galileo Galilei, Universit\` a di Padova, I-35131 Padova, Italy}
\affiliation{Istituto Nazionale di Fisica Nucleare Sezione di Padova, I-35131 Padova, Italy}
}

\newcommand\nichome{
\affiliation{Dipartimento di Fisica Galileo Galilei, Universit\` a di Padova, I-35131 Padova, Italy}
\affiliation{Istituto Nazionale di Fisica Nucleare Sezione di Padova, I-35131 Padova, Italy}
\affiliation{Istituto Nazionale di Astrofisica Osservatorio Astronomico di Padova, Italy}
}

\newcommand\kimhome{
\affiliation{Texas Center for Cosmology and Astroparticle Physics, Weinberg Institute, Department of Physics, The University of Texas at Austin, Austin, Texas 78712, USA}
}

\author{Jessie de Kruijf}
\email{jessiearnoldus.dekruijf@phd.unipd.it}
\jessiehome

\author{Eleonora Vanzan}
\email{eleonora.vanzan@phd.unipd.it}
\elehome

\author{Kimberly K.~Boddy}
\email{kboddy@physics.utexas.edu}
\kimhome

\author{Alvise Raccanelli}
\email{alvise.raccanelli.1@unipd.it}
\alvisehome

\author{Nicola Bartolo}
\email{nicola.bartolo@pd.infn.it}
\nichome

\begin{abstract}
The primordial power spectrum of curvature perturbations has been well-measured on large scales but remains fairly unconstrained at smaller scales, where significant deviations from $\Lambda$CDM may occur. Measurements of 21-cm intensity mapping in the dark ages promise to access very small scales that have yet to be probed, extending beyond the reach of cosmic microwave background and galaxy surveys. In this paper, we investigate how small-scale power-law enhancements---or blue tilts---of the primordial power spectrum affect the 21-cm power spectrum. We consider generic enhancements due to curvature modes, isocurvature modes, and runnings of the spectral tilt. We present forecasts for Earth- and lunar-based instruments to detect a blue-tilted primordial spectrum. We find that an Earth-based instrument capable of reaching the dark ages could detect any enhancements of power on nearly all the scales it can observe, which depends on the baseline of the interferometer. The smallest scales observed by such an instrument can only detect a very strong enhancement. However, an instrument on the far side of the Moon of the same size would be able to probe shallower slopes with higher precision. We forecast results for instruments with $100 \, {\rm km} \, (3000 \, {\rm km})$ baselines and find that they can probe up to scales of order $k_{\rm max} \sim 8 \, {\rm Mpc}^{-1} \, (k_{\rm max} \sim 250 \, {\rm Mpc}^{-1})$, thereby providing invaluable information on exotic physics and testing inflationary models on scales not otherwise accessible.
\end{abstract}

\maketitle


\section{Introduction}
\label{sec:intro}

The power spectrum of primordial perturbations describes the initial conditions of the Universe and thus of structure formation.
It is well-characterized by curvature perturbations $\zeta$ on relative scales from roughly $10^{-4}$ to $1 \, {\rm Mpc}^{-1}$, by measurements of cosmic microwave background (CMB) anisotropies and large-scale structure (LSS)~\cite{Planck2018VI, Kalaja2019,  Ivanov_2020,Philcox_2022, Ivanov_2021, DESI_2024_cosmological_parameters, Planck2018X}.

Previous work has constrained the primordial power spectrum on smaller scales using the abundance of ultracompact minihalos~\cite{Bringmann2012}, the distribution of stars in an ultrafaint dwarf galaxy~\cite{graham2024constraints}, spectral distortions of the CMB~\cite{Silk1981,Chluba2021}, and the high-redshift UV galaxy luminosity function~\cite{Sabti_2022, Yoshiura_2020}.
However, these methods can have large astrophysical uncertainties, preventing them from achieving the same level of sensitivity and robustness as the constraints from CMB anisotropies and LSS.
Possible future CMB experiments, such as PIXIE~\cite{kogut2024primordial} and CMB-HD~\cite{hdcollaboration2022snowmass2021cmbhdwhitepaper}, could reach scales up to $k\sim10 \, {\rm Mpc}^{-1}$.

Probing the primordial power spectrum on small scales could be the key to discovering new physics~\cite{Bechtol_2022, Boddy_2022}.
Many exotic physics models generate an increase (or blue tilt) of the small-scale power spectrum, such as nonstandard inflationary scenarios~\cite{Martin2001, Gong2011, Germani2017, Ezquiaga2018, Ballesteros2020, Mishra2020, Braglia2020, Wang2024, Mollerach1994, Garc_a_Bellido_1997, Barrow_1993, Palma_2020, Fumagalli_2023, ebadi2023gravitational, Kasuya_1997,  Wands_2002, Bartolo_2001, Linde_1997}, gravitational particle production of dark matter~\cite{Chung:2004nh, Chung:2011xd, Chung:2013rda, Ling:2021zlj, Kolb:2023ydq, Graham_2016, cyncynates2023detectabledefectfreedarkphoton}, axions~\cite{Chung:2016wvv, Chung:2023xcv, Cyncynates_2022, chung2024largebluespectralindex, Kasuya_2009}, quantum decoherence during inflation~\cite{Martin2018}, and bouncing cosmologies~\cite{Qiu_2011, Cai2012}.
Such blue-tilted primordial (or matter) power spectra have been investigated using CMB/LSS~\cite{Lucchin_1996, Chung_2018}, spectral distortions~\cite{Chluba2013}, dark matter substructure~\cite{Ando_2022, graham2024constraints, Josan_2010, Bringmann2012}, quasar light curves~\cite{karami2018forward}, Lyman-$\alpha$ forest~\cite{Stefano_1996,Afshordi_2003}, and Milky Way satellite velocities~\cite{esteban2023milky, dekker2024constraintsblueredtilted}.
There have also been forecasts for constraining blue tilts with Euclid and MegaMapper~\cite{chung2023}, as well as for Square Kilometer Array Observatory (SKAO) using 21-cm line-intensity-mapping (LIM) at relatively low redshifts $(z < 20)$~\cite{Sekiguchi_2014}.

In this work, we consider the sensitivity that future 21-cm fluctuation measurements from the dark ages have to enhancements of small-scale power, with respect to $\Lambda$CDM.
The main benefit of measuring the 21-cm power spectrum during the dark ages~\cite{Pritchard2011, Furlanetto2006, Loeb2003, Zaldarriaga_2004, Loeb_2004, Furlanetto_2006_global}, which spans the redshift range $30 \lesssim z \lesssim 200$, is that there is no astrophysical contamination, since the first stars have not yet formed; thus, we can more directly probe early Universe cosmology.
Moreover, fluctuations are not affected by Silk damping~\cite{Silk1968} and therefore remain undamped down to the baryon Jeans scale $k_J \sim 300 \, {\rm Mpc}^{-1}$.
As a result, we can access modes at much smaller scales with 21-cm LIM, compared to the CMB.

There are several proposals for low-frequency interferometers that would be able to detect the dark ages 21-cm signal.
Most are lunar-based experiments~\cite{DARE_2017, DAPPER_2019, FARSIDE_2019,  LUSEE2023, Hongmeng2023, FarView2024, DALI, Price_2018}, as well as a lunar-orbiting CubeSat~\cite{Artuc:2024rpq}.
It was also the aim of several Earth-based instruments to observe the dark ages~\cite{Bevins_2022, SARAS3_2021, Blyth2015AdvancingAW, maio2015bulkflowsenddark, MIST2024}.
While the current designs are not yet capable of reaching such high redshifts, future versions of them potentially could.
We consider the sensitivity to enhancements of small-scale power for a few hypothetical instruments, both Earth- and lunar-based.

Previous studies have investigated how changes to the matter power spectrum affect the 21-cm signal~\cite{Jones_2021, Vanzan:2023gui,Hotinli:2021vxg, Flitter:2022pzf, Munoz_2020,Munoz2017, Cole_2020,short2022dark, Driskell_2022, Ali-Haimoud2013}; most of them consider a suppression or enhancement of small-scale structure for particular models.
We take an agnostic approach by considering generic power-law enhancements of the primordial power spectrum due to a small-scale enhancement of curvature modes, the addition of uncorrelated cold dark matter isocurvature modes to the standard $\Lambda$CDM curvature modes, and a running of the spectral tilt.
We find that an Earth-based instrument can already probe most power spectra increases on the observable scales, while a lunar-based instrument of the same size can probe shallower slopes and obtain a higher accuracy.
When increasing the size of the instrument on the Moon, one can probe a much larger range of scales, allowing revolutionary and stringent tests of exotic physics and inflationary models.

This paper is organized as follows.
In Sec.~\ref{sec:blue-tilted power spectrum} we introduce different blue-tilted power spectra considered in this work. 
In Sec.~\ref{sec: 21cm dark ages} we give a brief overview of 21-cm LIM from the dark ages.
In Sec.~\ref{sec:method} we describe our methods for measuring the blue-tilted power spectrum and provide the instrument specifications we use for our forecasts.
In Sec.~\ref{sec:results} we show the results of our forecasts.
We summarize our findings and conclude in Sec.~\ref{sec: conclusion}.


\section{Blue-tilted primordial power spectrum}
\label{sec:blue-tilted power spectrum}

Within the standard $\Lambda$CDM scenario, the dimensionless primordial curvature power spectrum
\begin{equation}
  \Delta_{\zeta}^2(k) = A_s \left(\frac{k}{k_0}\right)^{n_s-1}
  \label{eq:P_adiabatic}
\end{equation}
is characterized by a tilt $n_s$ and amplitude $A_s$ at the pivot scale $k_0$.
Measurements of CMB anisotropy from the \textit{Planck} satellite, in conjunction with LSS data, show that this power spectrum is nearly scale invariant and red tilted (i.e.,~decreases with $k$)  with $n_s =  0.9649 \pm 0.0084$ ($95\%$ CL, \textit{Planck} TT, TE, EE + lowE + lensing), and has an amplitude given by $\ln(10^{10}A_s) = 3.044 \pm 0.028$ ($95\%$ CL, \textit{Planck} TT, TE, EE + lowE + lensing) at $k_0 = 0.05 \, {\rm Mpc}^{-1}$~\cite{Planck2018VI}.
These values are compatible with the simplest prediction of single-field, slow-roll inflation.

While the CMB and LSS set robust and stringent constrains on large scales, its behavior is largely unconstrained on small scales, allowing for the possibility of physics beyond $\Lambda$CDM to enter at $k > 1-10\, \rm{Mpc^{-1}}$.

Small-scale enhancements arise from a variety of early Universe physics, such as inflationary scenarios~\cite{Martin2001, Gong2011, Germani2017, Ezquiaga2018, Ballesteros2020, Mishra2020, Braglia2020, Wang2024, Mollerach1994, Garc_a_Bellido_1997, Barrow_1993, Palma_2020}, gravitational particle production of dark matter~\cite{Chung:2004nh, Chung:2011xd, Chung:2013rda, Ling:2021zlj, Kolb:2023ydq, Graham_2016}, axions~\cite{Chung:2016wvv, Chung:2023xcv}, quantum decoherence during inflation~\cite{Martin2018}, and bouncing cosmologies~\cite{Qiu_2011, Cai2012}.
These scenarios produce blue-tilted enhancements in the total primordial power spectrum $\Delta^2(k)$, either by directly increasing curvature power spectrum at small scales or by generating small-scale cold dark matter isocurvature.

In this section, we consider a primordial power spectrum that becomes blue-tilted (i.e.,~increases with $k$) at scales smaller than $1\, \rm{Mpc^{-1}}$.
We study enhancements of three different types: a curvature power spectrum with a broken power law; a blue-tilted cold dark matter isocurvature power spectrum; and the running of the spectral tilt, $n_s(k)$, which is a necessary feature of a primordial power spectrum that transitions from a red tilt at large scales to a blue tilt at small scales.

\subsection{Enhanced curvature power spectrum}
\label{subsec: enhanced curvature power spectrum}

We start by considering primordial blue-tilted curvature power spectra at small scales, which can be produced by e.g.~\cite{Mollerach1994, Martin2001, Gong2011, Germani2017, Ezquiaga2018, Ballesteros2020, Mishra2020, Wang2024, Garc_a_Bellido_1997, Barrow_1993, Palma_2020, Martin2018, Qiu_2011, ebadi2023gravitational, Fumagalli_2023}.
The primordial spectrum may possess additional features, such as a turnover that produces a bump at some large value of $k$.
Since such features are model dependent, we do not account for them.
However, we impose a threshold of $\Delta_{\zeta}^2(k) \leq 0.01$, forcing the power spectrum to remain constant at arbitrarily large $k$.
The presence of the threshold has no significant impact on our analyses.%
\footnote{There are cases in which the threshold is imposed at scales relevant to our analyses. However, these situations occur in the signal-dominated regime, so our projections based on signal-to-noise ratios remain insensitive to the threshold.}

\begin{figure}[t]
\includegraphics[width=.48\textwidth]{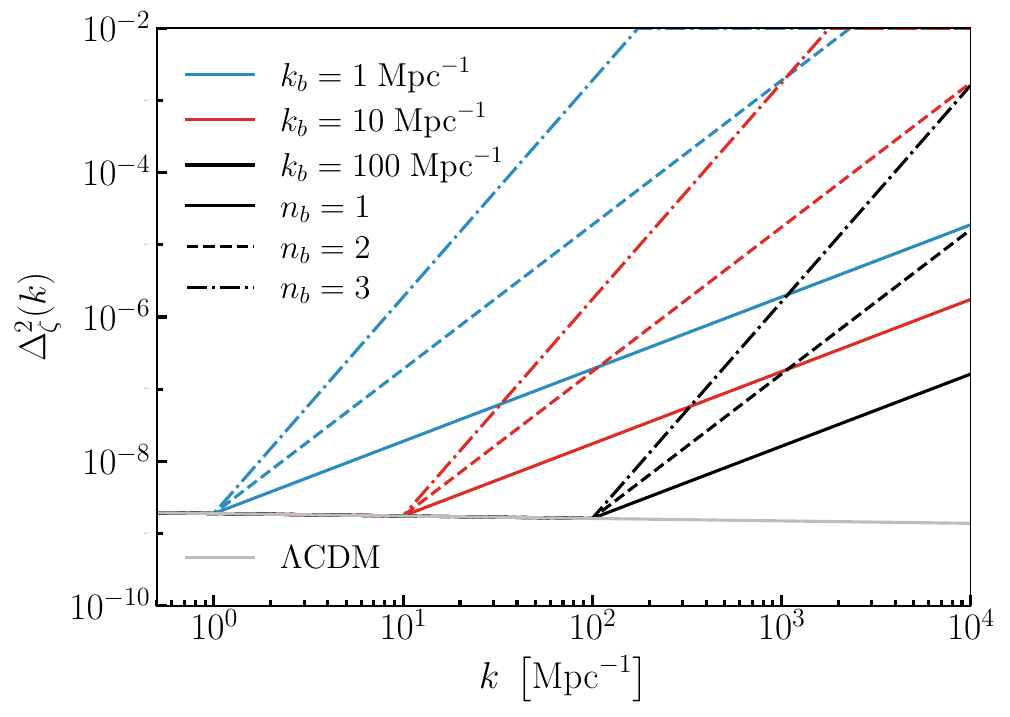}
  \caption{Dimensionless primordial curvature power spectrum.
    We show the power-law break from the $\Lambda$CDM case (solid gray) at various values of $k_{b}$, with a blue-tilted enhancement for $k>k_{b}$ for various values of the power-law index $n_b$.}
  \label{fig: primordial adiabatic}
\end{figure}

To investigate small-scale enhancements in a model-independent manner, we adopt a simple broken-power-law parametrization of the dimensionless primordial curvature power spectrum:
\begin{equation}
  \Delta^2_\zeta (k) =
  \begin{cases}
    A_s \left( \dfrac{k}{\kpivot} \right)^{n_s-1} \qquad & k < k_{b} \\
    A_b \left( \dfrac{k}{k_{b}} \right)^{n_b} & k > k_{b} \, ,
  \end{cases}
  \label{eq:broken}
\end{equation}
where $k_b$ is the scale where the power spectrum breaks and begins to increase as a power law with index $n_b$.
The amplitude $A_b$ of the small-scale power is defined at the scale $k_b$, and continuity demands $A_b = A_s \left( k_{b} / \kpivot \right)^{n_s-1}$.
Assuming $k_{b} \gtrsim 1\, {\rm Mpc}^{-1}$, $\Delta^2_\zeta (k)$ coincides with Eq.~\eqref{eq:P_adiabatic} for $k < k_{b}$ and thus remains within the limits set by CMB and LSS on large scales.
On smaller scales $k > k_{b}$, we allow for new physics to produce a blue-tilted spectrum. Figure~\ref{fig: primordial adiabatic} illustrates the power-law break in Eq.~\eqref{eq:broken} for various values of $k_{b}$ and $n_b$.

The power-law scaling for $k>k_{b}$ in Eq.~\eqref{eq:broken} can be mapped to various models.
For example, for tilted hybrid inflation~\cite{Garc_a_Bellido_1997}, the spectral index and the amplitude of the power spectrum can be mapped to the effective mass of the inflaton field and the value of the field at the moment when perturbations that have momentum $k$ at the end of inflation were produced.
Another example is quantum decoherence during inflation~\cite{Martin2018}. The evolution of inflationary curvature perturbations can be modified due to the interaction with other degrees of freedom present in the early Universe.
This interaction leads to quantum decoherence, and in various cases to a blue-tilted correction of the power spectrum on scales considered in this work.
The time dependence and strength of the interaction between the perturbations and other degrees of freedom determine $n_b$ and $k_b$. 

\subsection{Enhanced dark matter isocurvature power spectrum}
\label{sec:iso}

\begin{figure}[t]
  \includegraphics[width=.48\textwidth]{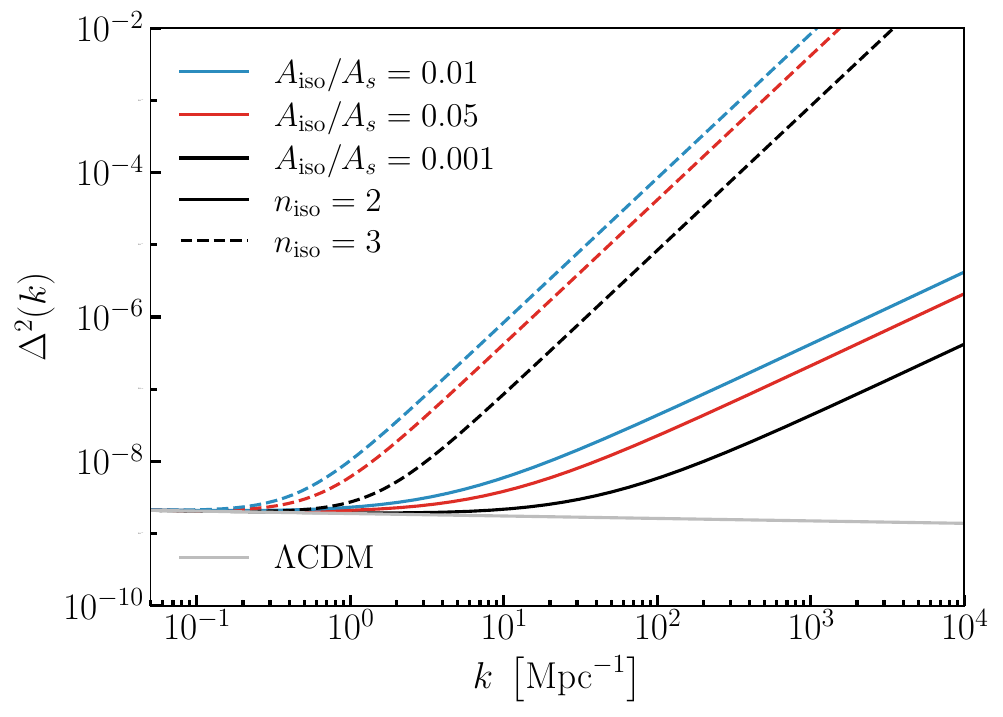}
  \caption{Total dimensionless primordial power spectrum $\Delta^2(k)$, for curvature and cold dark matter isocurvature modes. 
    We show the increase of the power spectrum, along with the $\Lambda$CDM case (solid gray), due to isocurvature modes for various values of $n_{\rm iso}$ and $ A_{\rm iso}/A_s$.}
  \label{fig: primordial isocurvature}
\end{figure}

A different class of small-scale enhancements arises from the production of isocurvature modes~\cite{Bartolo_2001, Linde_1997, Ghosh_2022,  Chung:2016wvv, Sekiguchi_2014, Kawasaki_2011, Kasuya_2009, Braglia2020, Chung:2004nh, Chung:2013rda, Graham_2016, Ling:2021zlj, Kolb:2023ydq, Kasuya_1997, Chung:2011xd, Chung:2023xcv, Wands_2002, chung2024largebluespectralindex}.
We assume that adiabatic modes give rise to the dimensionless primordial curvature power spectrum in Eq.~\eqref{eq:P_adiabatic}, and the isocurvature modes produce a dimensionless primordial isocurvature power spectrum
\begin{equation}\label{eq:iso PS}
  \Delta_{\mathcal{S_{\rm cdm}}}^2(k) = A_{\rm iso} \left(\frac{k}{\kpivot}\right)^{n_{\rm iso}-1} \, ,
\end{equation}
with a tilt $n_{\rm iso}$ and an amplitude $A_{\rm iso}$ at the \textit{Planck} pivot scale $\kpivot$.
Although the adiabatic and isocurvature modes are generically correlated, we assume for simplicity that they have no cross-correlation power, such that the total dimensionless primordial power spectrum is 
\begin{equation}\label{eq: full PS iso}
  \Delta^2(k) = \Delta_{\zeta}^2(k) + \Delta_{\mathcal{S_{\rm cdm}}}^2(k) \, ,
\end{equation}
as demonstrated in Fig.~\ref{fig: primordial isocurvature}. 
For all scenarios considered in this work, we have checked explicitly that the amplitudes are consistent with constraints from \textit{Planck}~\cite{Planck2018X}.
Similar to the curvature case, we impose a threshold of $\Delta^2(k) \leq 0.01$.

A blue-tilted isocurvature power spectrum can arise, for example, from the QCD axion~\cite{Chung:2023xcv, Kasuya_2009, Chung_2022_analytic, Kasuya_1997}, which produces an isocurvature spectral index of $1<n_{\mathrm{iso}}\leq 4$.
The bounds on the isocurvature parameters can be translated to a bound on the inflationary energy scale $H_{I}$~\cite{Hamann2009, Hertzberg2008}.
A similar translation can be done in the context of ultralight axions~\cite{Marsh2013}, for which $H_{I}$ depends on $A_{\rm iso}/A_s$ and the fraction of dark matter that ultralight axions constitute.

Alternatively, a blue tilt with $n_{\rm iso}=4$ can be generated by particle production mechanisms with a finite correlation length. The resulting isocurvature power spectrum vanishes on scales smaller than a cutoff scale determined by causality, and only white noise is left on these smaller scales.
Examples of such mechanisms are postinflationary dark matter production~\cite{amin2024lowerbounddarkmatter, Caprini_2009}, vector dark matter~\cite{Graham_2016, Gorghetto_2022}, and Peccei Quinn breaking during inflation with subsequent axion production~\cite{Redi_2022}.

\subsection{Running parameters}
\label{sec:runnings}

The parametrization in Eq.~\eqref{eq:broken} is meant to describe the main features of the primordial spectrum; however, if there is a transition from the decreasing to the increasing behavior as a function of $k$, this transition would be described by the first and higher order derivatives of the spectrum.
We parametrize the primordial power spectrum in this case as~\cite{Kosowsky_1995}
\begin{equation}
\begin{split}
  \Delta_{\zeta}^2(k) =& A_s \left( \frac{k}{\kpivot} \right)^{n_s-1} \\
  & \times \exp \left[ \frac{1}{2}\alpha_s \log^2\left(\frac{k}{\kpivot}\right) +\frac{1}{6}\beta_s \log^3\left(\frac{k}{\kpivot}\right) \right] \, ,
  \label{eq:runnings}
\end{split}
\end{equation}
where
\begin{equation}
  \alpha_s = \frac{d n_s}{d \log k} \, , \qquad \beta_s = \frac{d \alpha_s}{d \log k} \, ,
\end{equation}
are the first and second running parameters of the tilt, respectively~\cite{Kosowsky_1995, Baumann_2009}.
\textit{Planck} 2018 data constrains the running parameters to be $\alpha_s = 0.013 \pm 0.012$ and $\beta_s = 0.022 \pm 0.012$ (at the pivot scale $k=0.05 \, \rm{Mpc}^{-1}$)~\cite{Planck2018VI, Cabass_2016, Planck2018X}.

Standard single-field slow-roll inflation predicts $\alpha_s \sim (1-n_s)^2 \sim \mathcal{O}(\epsilon^2,\eta^2) \sim 10^{-3}$ and $\beta_s \sim (1-n_s)^3 \sim \mathcal{O}(\epsilon^3, \eta^3) \sim 10^{-5}$~\cite{Adshead:2010mc, Munoz2017, Kosowsky_1995, Lidsey_1997}.
Therefore, high precision constraints on the running parameters could either confirm or rule out this inflationary scenario and permit tests of other inflationary models that predict a running~\cite{Easther_2022, Bahr_Kalus_2023, Kohri_2013, Munoz2017}, such as multifield~\cite{Gong_2015}, and warm inflation~\cite{Das_2023}.

Note that this parametrization is only valid for small deviations from the scale-invariant case and is thus not appropriate for the large enhancements studied in Sec.~\ref{subsec: enhanced curvature power spectrum}.
We focus on the region of $k$ in which the power spectrum begins to transition to becoming blue tilted, and thus the parametrization is valid.


\section{21-cm intensity mapping during the dark ages}
\label{sec: 21cm dark ages}

In this section we describe the main observable of interest: the intensity mapping signal of 21-cm neutral hydrogen.
We provide a brief review of key concepts for 21-cm LIM from the dark ages, before the formation of the first stars, which occurs around $z\sim 30$ in $\Lambda$CDM.%
\footnote{We note that enhancements of small-scale structure should lead to advanced star formation. Studying the onset of cosmic dawn is beyond the scope of this work, and we use $z=30$ as a convenient reference redshift for our forecasts. We assume no Lyman-$\alpha$ coupling throughout the entire dark ages, including $z=30$.}
For more details, we refer the reader to Refs.~\cite{Madau1996,Zaldarriaga_2004,Pritchard2008,Pritchard2010,Furlanetto2019} for 21-cm LIM, Refs.~\cite{Loeb2003,Furlanetto2006,Lewis2007,Furlanetto2009,Pritchard2011, Iliev_2002} for specific application to the dark ages, and Refs.~\cite{Kovetz2017,Bernal2022} for recent reviews on LIM.

Post recombination, the Universe is filled with neutral hydrogen gas clouds.
The CMB photons act as a backlight, scattering on neutral hydrogen atoms and exciting the 21-cm hyperfine transition between the singlet and the triplet state.
During the dark ages, the temperature of the gas is lower than that of the CMB, and there is a net absorption of CMB photons of wavelength $\lambda_{21} \approx 21\, {\rm cm}$ with corresponding frequency $\nu_{21} \approx 1420 \, {\rm MHz}$.
The remaining CMB photons at the transition frequency are redshifted as the Universe expands.
Therefore, observing a deficit of CMB photons at a given frequency identifies a unique redshift slice for the absorption process, thereby enabling tomographic analyses.

The observed quantity for the 21-cm signal is the brightness temperature contrast with respect to CMB~\cite{Ali-Haimoud2013},
\begin{equation}
  \label{eq:T21}
  T_{21} = \frac{T_s-T_{\rm CMB}}{1+z} (1-e^{-\tau}) \approx \tau \frac{T_s-T_{\rm CMB}}{1+z} \, ,
\end{equation}
where we have assumed the optically thin limit in which the optical depth $\tau \ll 1$.
The spin temperature $T_s$ is defined by the ratio of abundances of neutral hydrogen in the triplet state $n_1$ and in the singlet state $n_0$:
\begin{equation}
  \frac{n_1}{n_0} \equiv 3 e^{-E_{10} / T_s} \approx 3\left(1-\frac{E_{10}}{T_s}\right) \, ,
\end{equation}
where $E_{10} \approx 5.9 \, \mu {\rm eV}$ is the energy difference between the two states.
Collisions of neutral hydrogen with other atoms and electrons drive $T_s \xrightarrow{} T_{\rm gas}$, while radiative transitions involving absorption of and emission to the radio background drive $T_s \xrightarrow{} T_{\rm CMB}$.
A nonzero $T_{21}$ is possible only after $T_{\rm gas}$ decouples from $T_{\rm CMB}$ at redshift $z \sim 200$.
Frequent collisions keep $T_s$ coupled to $T_{\rm gas}$, allowing for a net absorption of 21-cm CMB photons by neutral hydrogen.
Around $z \sim 100$, the expansion of the Universe dilutes the gas sufficiently to render collisions inefficient, and $T_s$ tends toward $T_{\rm CMB}$.

The Sobolev optical depth~\cite{Sobolev_1957} is 
\begin{equation}
  \tau = \frac{3}{32\pi} \frac{E_{10}}{T_s} x_{\rm HI} n_H \lambda_{21}^3 \frac{A_{10}}{H(z) +(1+z) \partial_r v} \, ,
\end{equation}
where $x_{\rm HI}$ is the fraction of neutral hydrogen, $n_H$ is the number density of hydrogen, $A_{10} \approx 2.85 \times 10^{-15} \, {\rm s}^{-1}$ is the spontaneous decay rate of the spin-flip transition, $H(z)$ is the Hubble parameter, and $\partial_r v$ is the line-of-sight gradient of the peculiar velocity of the gas.
During the dark ages, $\tau \ll 1$, and the approximation in Eq.~\eqref{eq:T21} is valid.

The brightness temperature is a function of the local hydrogen density and the gas temperature, and it can be parametrized as~\cite{Ali-Haimoud2013}
\begin{equation}
  T_{21} = \Bar{T}_{21} \left( 1+\delta_v \right) +c_b\delta_b +c_T\delta_{T_{\rm gas}} \, ,
\end{equation}
where $\delta_b=\delta n_b/\Bar{n}_b \simeq \delta_H$ up to negligible corrections, $\delta_{T_{\rm gas}} = \delta T_{\rm gas} / \Bar{T}_{\rm gas}$, and $\delta_v = -(1+z)\partial_rv/H(z)$.
The coefficients $c_{b,T}$ are functions of redshift only; the mean brightness temperature $\Bar{T}_{21}$ is defined by setting all perturbations to zero.
Following Ref.~\cite{Short2019}, we neglect fluctuations in the gas temperature and in the neutral hydrogen fraction to write the 21-cm brightness temperature fluctuations as
\begin{equation}
  \label{eq:T21fluctuations}
  \delta T_{21}(\vec{x},z) \simeq \alpha(z) \delta_b(\vec{x},z) +\Bar{T}_{21}(z) \delta_v(\vec{x},z) \, ,
\end{equation}
where $\alpha(z) = dT_{21} / d\delta_b$.

We expand the angular dependence of the brightness fluctuations in the basis of spherical harmonics,
\begin{equation}
  \delta T_{21}(\vec{x},z) = \sum_{\ell m} a_{\ell m}(z) Y_{\ell m}(\vhat{x}) \, ,
\end{equation}
and write the angular power spectrum of 21-cm fluctuations as
\begin{equation}
  \langle a_{\ell m}(z) a_{\ell' m'}(z) \rangle = C_{\ell}(z) \delta_{\ell \ell'} \delta_{m m'} \, .
\end{equation}
By combining Eq.~\eqref{eq:T21fluctuations} with the continuity equation, we can directly relate the 21-cm angular power spectrum to the matter power spectrum~$P_m(k)$~\cite{Bharadwaj2004,Ali-Haimoud2013,Short2019}:
\begin{equation}
  C_{\ell}(z) = \frac{2}{\pi} \int_0^{\infty} k^2 dk\, \mathcal{T}_{\ell}(k,z)^2 P_m(k) \, ,
\end{equation}
with 
\begin{equation}
  \hspace{-0.2cm} \mathcal{T}_{\ell}(k,z)\hspace{-0.1cm} = \hspace{-0.2cm}\int_0^{\infty} \hspace{-0.4cm} dx \, W_{\nu}(x) \hspace{-0.1cm} \left[ \alpha(z) j_{\ell}(kx) -\bar{T}_{21}(z) \frac{\partial^2 j_{\ell}(kx)}{\partial (kx)^2} \right] ,
  \label{eq:21cm-transfer}
\end{equation}
where $W_\nu (x)$ is a window function and $j_\ell (kx)$ is the spherical Bessel function of the first kind.

The matter power spectrum is linked to the primordial curvature power spectrum via
\begin{equation}
  \label{eq: matter PS}
  P_m(k,z) = \sum_X D_X(z)^2 \frac{2\pi^2}{k^3} \Delta^2_X(k) T_{m,X}(k)^2 \, ,
\end{equation}
where $X=\{\zeta, \mathcal{S}_{\rm cdm}\}$ represents the initial condition, and we assume the adiabatic and cold dark matter isocurvature modes are uncorrelated.
We use \texttt{CLASS}~\cite{Diego_Blas_2011} to compute the linear growth factor $D_X(z)$ and the matter transfer function $T_{m,X}(k)$.
In the $\Lambda$CDM case, the power spectrum is still linear for the scales and redshifts we probe.
In the cases where the primordial power spectrum reaches values above $10^{-3}$ at the scales of interest, the resulting matter power spectrum could become nonlinear.
For our results, the majority of the sensitivity is near the $\ell_{\rm {max}}$ value (see Sec.~\ref{sec:results} for more details), and we do not expect our conclusions to be affected by assuming the matter power spectrum is still linear; a careful investigation of this issue is left for future work.

The window function $W_{\nu}(x)$ in Eq.~\eqref{eq:21cm-transfer} accounts for the finite spectral resolution of the instrument.
We model the window function as a tophat, centered at the radial distance $x(z)$ corresponding to the targeted redshifted frequency $\nu=\nu_0/(1+z)$.
The width of the window
\begin{equation}
  \Delta x = \frac{(1+z)^2}{\nu_0 H(z)} B
\end{equation}
depends on the frequency bandwidth $B$.


\section{Methodology}
\label{sec:method}

In this section we describe our procedure for forecasting the sensitivity of future 21-cm LIM experiments to enhancements of small-scale power.
We consider three proposed future instruments as examples: one Earth-based instrument, inspired by the Square Kilometre Array Observatory, and two lunar instruments with different sensitivities.

For our forecasts, we take $\Lambda$CDM as a reference model, using the \textit{Planck} 2018 best fit values~\cite{Planck2018VI} for the fiducial cosmological parameters: $\{ \omega_b=0.02238, \  \omega_c=0.1201, \  h=0.6781, \  n_s=0.96605, \  \ln 10^{10} A_s=3.0448  \}$.

\subsection{Forecasting setup}
\label{sec:forecasts}

We estimate the detectability of a given set of parameters controlling the deviation from $\Lambda$CDM of the primordial power spectrum by calculating the signal-to-noise ratio (SNR) as~\cite{Scelfo2018}
\begin{equation}
  \text{SNR}^2 \approx f_{\rm sky} \sum_z \sum_{\ell} \left( \frac{C_{\ell}^{\rm blue}(z) -C_{\ell}^{\Lambda \rm CDM}(z)}{\sigma_{\ell}} \right)^2 \, ,
  \label{eq:SNR}
\end{equation}
where $C_{\ell}^{\rm blue}$ and $C_{\ell}^{\Lambda \rm CDM}$ are the blue-tilted and $\Lambda$CDM 21-cm angular power spectra, respectively, $f_{\rm sky}$ is the fraction of the sky observed, and the variance is
\begin{equation}
  \sigma_{\ell}^2 = \frac{2\left(C_{\ell}^{\Lambda\rm CDM}+C_{\ell}^{\rm noise}\right)^2}{(2\ell+1)} \, .
\end{equation}
The sum in Eq.~\eqref{eq:SNR} runs over all redshift bins and over all multipoles up to the maximum observable multipole
\begin{equation}
  \label{eq:lcover}
  \ell_{\rm max}(z) = 2\pi \frac{D_{\rm base}}{\lambda(z)} \, ,
\end{equation}
where $D_{\rm base}$ the baseline of a given instrument, and $\lambda(z)= \lambda_{21} (1+z)$ is the redshifted 21-cm wavelength.

We model the noise power spectrum as~\cite{Shiraishi2016}
\begin{equation}
  \label{eq:noise}
  C_{\ell}^{\rm noise}(z) = (2\pi)^3 \frac{T_{\rm sys}^2(\nu)}{B \,  t_{\rm obs} f_{\rm cover}^2} \left(\frac{1}{\ell_{\rm max}(z)}\right)^2 \, ,
\end{equation}
where $t_{\rm obs}$ is the total time of observation, and $f_{\rm cover}$ is the coverage fraction (i.e.,~the fraction of the array area that is covered by antennas). We emphasize Eq.~\eqref{eq:noise} is the noise power spectrum of the instrument. In this work we do not consider possible additional sources of noise such as foregrounds, lensing, or atmospheric noise for the Earth-based instrument.
We take the system temperature to be the synchrotron temperature of the observed sky~\cite{Shiraishi2016}:
\begin{equation}
  T_{\rm sys}(\nu) = 180 \left(\frac{180\text{ MHz}}{\nu}\right)^{2.6} \, .
\end{equation}

We assume that the redshift bins are independent, which is a good approximation provided that the correlation length in frequency is smaller than the bandwidth of the instrument.
We define the radial correlation length $\xi_r$ as the radial separation, beyond which the cross-correlation between two redshift slices is less than 1/2 of the power spectrum~\cite{Munoz2015}.
The corresponding correlation length in frequency is
\begin{equation}
  \xi_{\nu} = \frac{d\nu}{dz} \frac{dz}{dr} \xi_r \approx 1\, \text{MHz} \left(\frac{51}{1+z}\right)^{1/2} \left(\frac{\xi_r}{60\, \text{Mpc}}\right) \, .
\end{equation}
We take linearly spaced bins in frequency, which we convert into redshift bins, and verify that $\xi_{\nu}$ is smaller than the bandwith $B$ of the instrument.
We neglect cross-correlations between different redshift bins, since they are expected to be negligible for 21-cm LIM measurements~\cite{Hall2012}.

For the case of running of the spectral tilt, we perform a Fisher matrix analysis~\cite{Fisher1935, coe2009fisher, Bellomo2020, Bernal2020, Verde_2010, heavens2010statisticaltechniquescosmology, Tegmark_1997, Vogeley_1996, Tegmark_2004, Seo_2003}, since we have a specific model with fiducial values for the running parameters ($\alpha_s$, $\beta_s$); whereas for the blue-tilted curvature and isocurvature scenarios, we use the SNR to investigate which values of $\{k_b, n_s\}$ or $\{A_{\rm iso},n_{\rm iso}\}$ we can test.
The Fisher approach approximates the likelihood around its maximum as a Gaussian and returns the smallest possible error on the parameter, namely the one set by the Cramer-Rao bound.
The Fisher information matrix is
\begin{equation}
  F_{\alpha\beta} = f_{\rm sky} \sum_{z} \sum_{\ell} \sigma_{\ell}^{-2} \frac{\partial C_{\ell}}{\partial\vartheta_{\alpha}} \frac{\partial C_{\ell}}{\partial\vartheta_{\beta}} \, ,
\end{equation}
where $\vartheta_{\alpha,\beta}$ represent our running parameters, and the sums run over all multipoles up to $\ell_{\rm max}$ and over all redshift bins, assuming there is no overlap in the amount of information contained in each bin.

\subsection{Instruments}
\label{sec:instruments}

There are several proposals for low-frequency interferometers that would be able to detect the redshifted 21-cm signal from the dark ages.
Most proposals are for lunar-based experiments, see e.g.,~\cite{DARE_2017, DAPPER_2019, FARSIDE_2019, LUSEE2023, Hongmeng2023,FarView2024, DALI, Price_2018, goel2022probingcosmicdarkages}, including lunar-orbiting CubeSats~\cite{Artuc:2024rpq}; more proposals are being developed for both lunar- and space-based interferometers~\cite{RaccanelliMoon:prep}.
These experiments are generally lunar-based, because the frequency of the redshifted signal from the end of the dark ages lies at the edge of where the Earth's ionosphere becomes opaque~\cite{Jester_2009, datta2016effectsionospheregroundbaseddetection}.
Moreover, the far side of the Moon should be completely radio quiet and stable~\cite{Koopmans2021}, making it an ideal place for such an instrument.

There is an open debate on how well we will be able to model the signal and actually observe the end of the dark ages from the (Earth) ground.
Several Earth-based instruments had planned to observe the dark ages~\cite{Bevins_2022, SARAS3_2021, Blyth2015AdvancingAW, maio2015bulkflowsenddark, MIST2024}, but this possibility is still uncertain; therefore, in this work we also consider an Earth-based instrument, assuming it will observe the very end of the dark ages, which we take to be at redshift $z=30$. We consider an idealized scenario where the signal is only affected by the noise of the instrument itself; the modeling of the noise from the Earth's atmosphere is beyond the scope of this work.

In order to keep our findings as general as possible, we consider three surveys, the details of which are presented in Table~\ref{tab:survey_specs}.
The instruments are an idealized future Earth-based instrument---chosen to be of similar size as SKAO, which we therefore refer to as SKAO-like, limited to $z=30$---and two possible configurations of a lunar radio array (LRA) on the far side of the Moon.
We refer to the lunar instruments as LRAI and LRAII; we consider a baseline of 100 km for the former, while the latter has a baseline that spans nearly the diameter of the Moon to demonstrate the maximal capabilities of a lunar array.
We also consider how our results are affected by changes to the total observation time $t_{\rm obs}$ and the coverage fraction $f_{\rm cover}$ in Appendix~\ref{sec:Appendix}. The specific distribution of the radio arrays can also affect the sensitivity of the instrument~\cite{Tegmark_2010, Wijnholds_2011}.

\begin{table}[t]
  \centering
  \begin{tabular}{|ll|C{1.6cm}|C{1.6cm}|C{1.6cm}|}
    \hline
    & &  SKAO-like & LRAI & LRAII \\
    \hline
    $B$ & [MHz]  & 2 & 2 & 2 \\
    \hline
    $D_{\rm base}$ & [km]  & 100 & 100 & 3000 \\
    \hline
    $f_{\rm cover}$ &  & 0.2 & 0.5 & 0.75 \\
    \hline
    $t_{\rm obs}$ & [yrs] & 10 & 10 & 10 \\
    \hline
    $f_{\rm sky}$ & & 0.75 & 0.75 & 0.75 \\
    \hline
  \end{tabular}
  \caption{Instrument specifications. For each configuration, we list the assumed bandwidth in frequency $B$, baseline $D_{\rm base}$, fraction of the instrument's total area that is covered by antennas $f_{\rm cover}$, observation time in years $t_{\rm obs}$, and sky coverage fraction $f_{\rm sky}$.}
  \label{tab:survey_specs}
\end{table}

\begin{figure}[t]
  \centering
  \includegraphics[width=.48\textwidth]{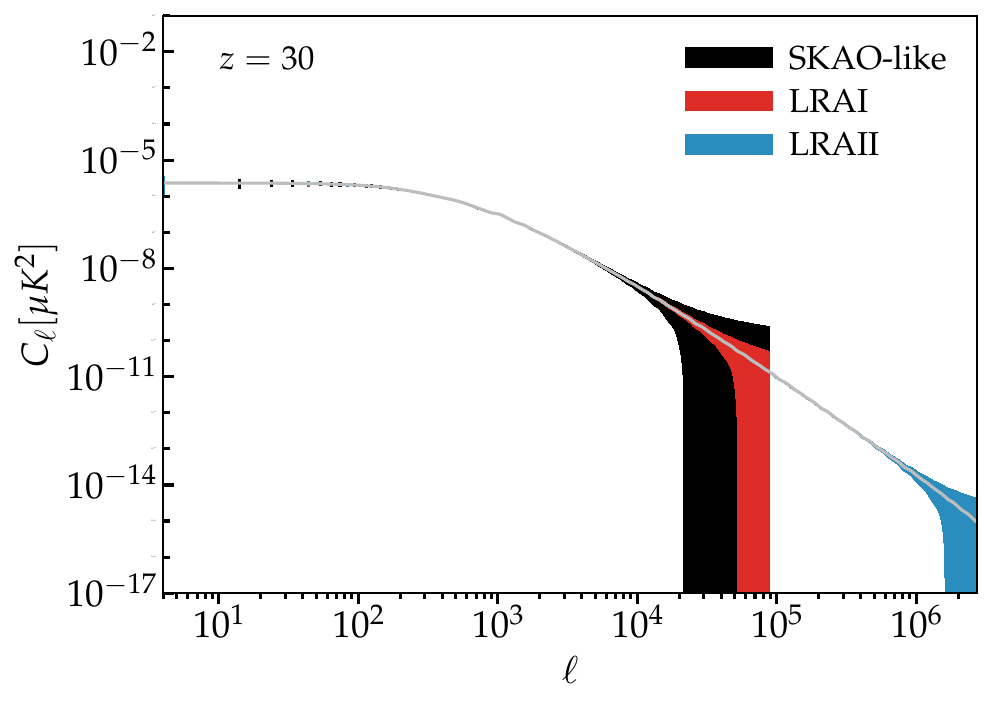}
  \includegraphics[width=.48\textwidth]{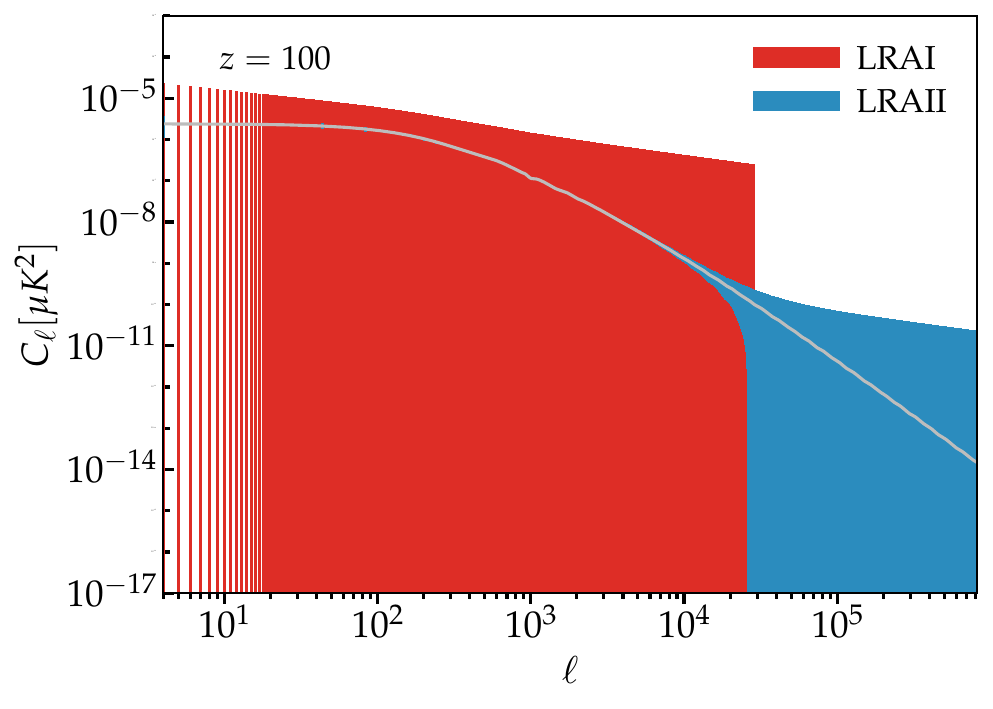}
  \caption{Angular power spectrum of 21-cm brightness temperature fluctuations at $z=30$ (top panel) and $z=100$ (bottom panel), with error bars for the instruments listed in Table~\ref{tab:survey_specs}.
    The theoretical prediction for $C_{\ell}^{\Lambda \rm CDM}$ is shown in gray for reference.
    The sharp cutoff at high multipoles corresponds to the maximum observable multipole.}
  \label{fig:Cls_with_error}
\end{figure}

For the instruments considered in this work, the predicted ($\Lambda$CDM) power spectrum and related uncertainties are shown in Fig.~\ref{fig:Cls_with_error} for $z=30$ (top) and $z=100$ (bottom).
The theoretical prediction for $C_{\ell}^{\Lambda \rm{CDM}}$ is shown in gray for reference. 
The largest multipole observable is directly related to the baseline, while the magnitude of the error bars is mostly influenced by the signal and the $f_{\rm{cover}}$.

\section{Results}
\label{sec:results}

Here we present our forecasted constraints on the models presented in Sec.~\ref{sec:blue-tilted power spectrum}, using the methodology of Sec.~\ref{sec:forecasts} for the instruments of Sec.~\ref{sec:instruments}.

\subsection{Enhanced curvature power spectrum}
\label{subsec:results curvature}

\begin{figure*}[!ht]
  \includegraphics[width=.99\textwidth]{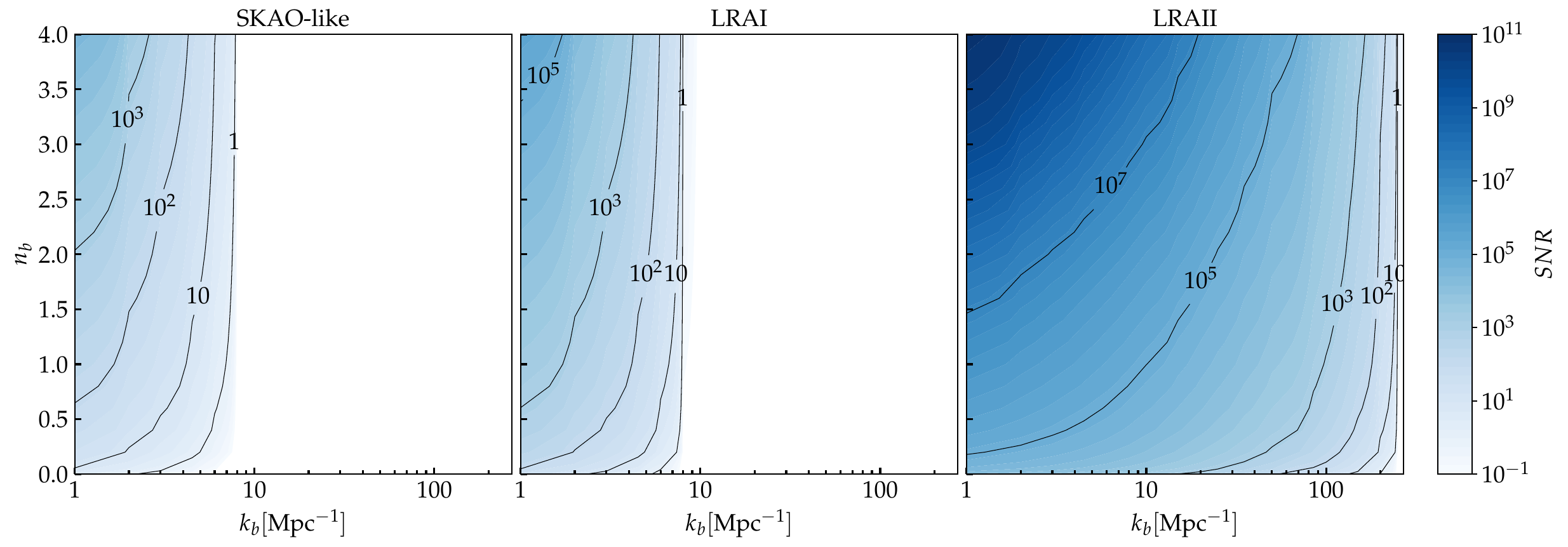}
  \caption{SNR of the blue-tilted deviation from $\Lambda$CDM for each of the instruments listed in Table~\ref{tab:survey_specs}, shown for varying values of $k_b$ and $n_b$.
    The white regions are those for which SNR $< 1$ and the instrument would have no sensitivity.
    The contour lines indicate several specific values of the SNR.}
  \label{fig:SNR kcut vs nb}
\end{figure*}

\begin{figure}
  \includegraphics[width=.49\textwidth]{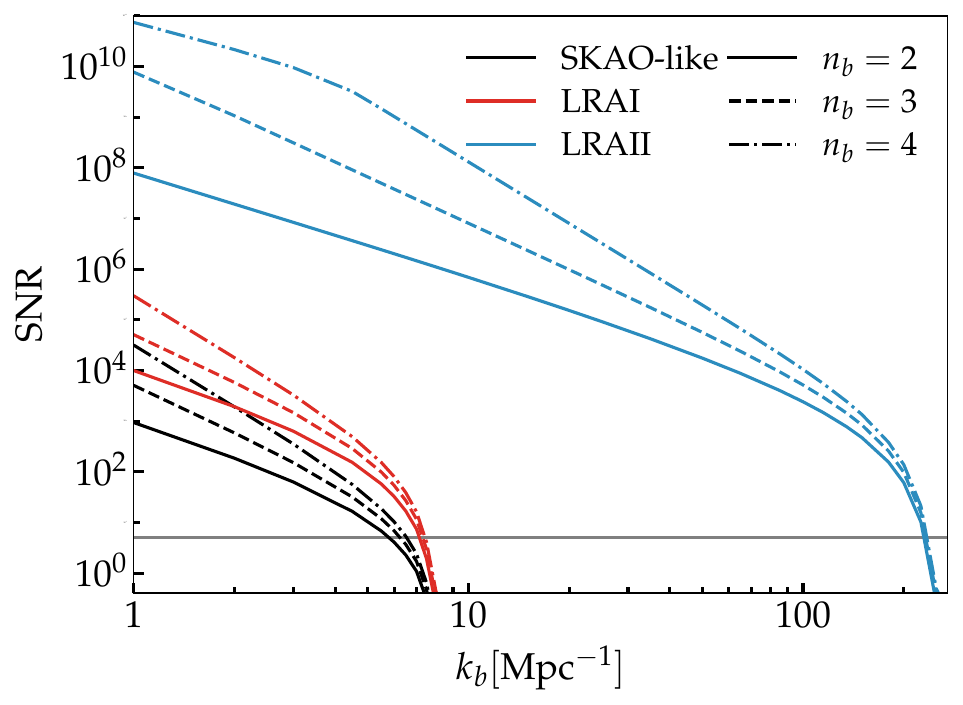}
  \caption{SNR of the blue-tilted deviation from $\Lambda$CDM for each of the instruments listed in Table~\ref{tab:survey_specs}, shown as a function of $k_{b}$, for $n_b =2,3,4$.
    We indicate $\rm{SNR}=5$ with a gray horizontal line for reference.}
  \label{fig:SNR different surveys}
\end{figure}

We start by showing results for the parametrization in Eq.~\eqref{eq:broken}.
In Fig.~\ref{fig:SNR kcut vs nb} we show the SNR for the detection of deviations from $\Lambda$CDM, as a function of the cutoff scale and slope $\{k_b, n_b\}$, for each instrument considered in this work.
In the $k$ range that is accessible to a given instrument, virtually any blue tilts would be easily detected with a high SNR.
We expect this result, given the error bars shown in Fig.~\ref{fig:Cls_with_error}.
Importantly, the baseline of the interferometer is a key feature for testing small-scale enhancements of the power spectrum.
This point is also illustrated in Fig.~\ref{fig:SNR different surveys}, where we show how the SNR changes as a function of $k_b$ for specific values of $n_b = 2,3,4$.

The highest value of $k_b$ that can be probed is nearly identical to the highest $k$ measured by the instrument, $k_{\rm max} = \ell_{\rm max} / \chi(z)$, where $\ell_{\rm max}$ is given in Eq.~\eqref{eq:lcover} and $\chi(z)$ is the comoving distance.
At $z=30$, $k_{\rm max} \sim 8 \, {\rm Mpc}^{-1}$ for the small instruments (SKAO-like, LRAI), while $k_{\rm max} \sim 230 \, {\rm Mpc}^{-1}$ for the larger instrument (LRAII).
Only for the lowest values of $n_b$ and the highest $k_b$ is the signal indistinguishable from the noise.

\subsection{Enhanced dark matter isocurvature power spectrum}
\label{subsec: results DM isocurvature}

\begin{figure*}
  \includegraphics[width=.99\textwidth]{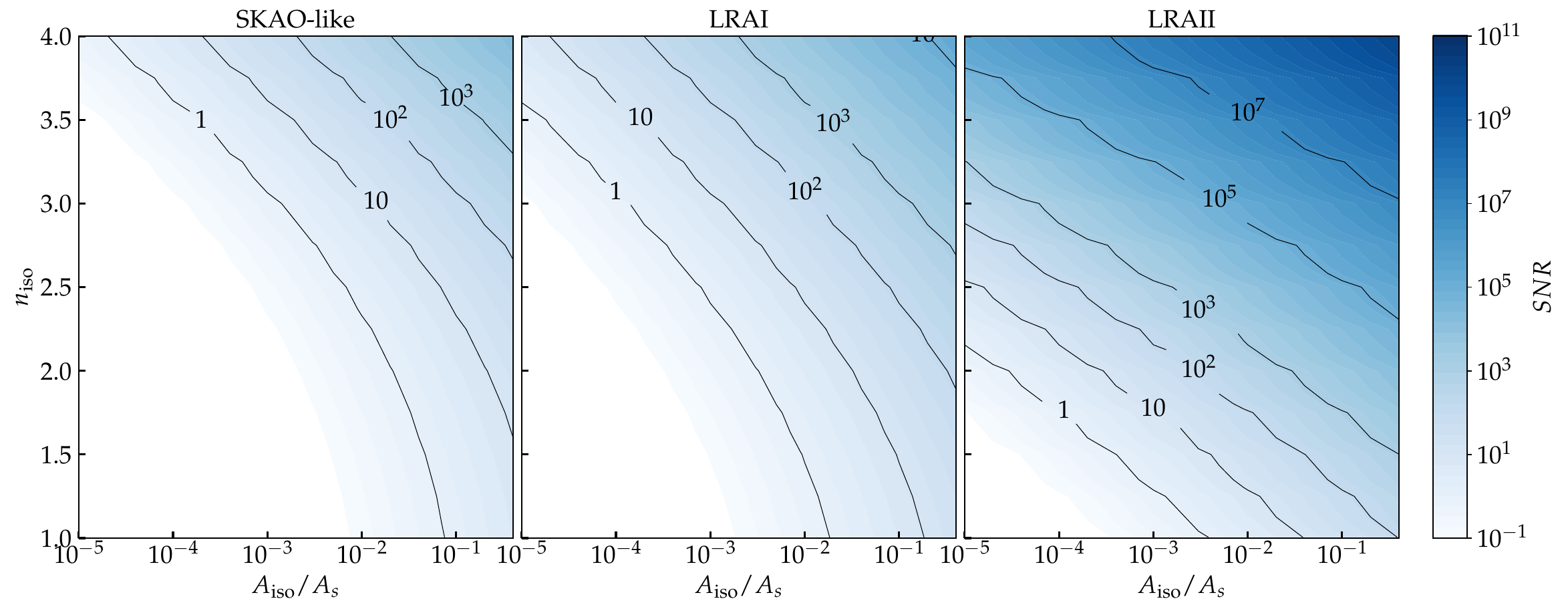}
  \caption{SNR for detecting blue-tilted cold dark matter isocurvature spectra, for each of the instruments listed in Table~\ref{tab:survey_specs}, shown for varying values of $A_{\rm iso}/A_s$ and $n_{\rm iso}$.
    The white regions are those for which SNR $< 1$ and the instrument would have no sensitivity.
    The contour lines indicate several specific values of the SNR.}
  \label{fig:SNR iso}
\end{figure*}

We now consider cold dark matter isocurvature modes that produce a blue-tilted isocurvature primordial power spectrum, given by Eq.~\eqref{eq:iso PS}.
Figure~\ref{fig:SNR iso} shows the SNR for the detection of isocurvature modes, as a function of the amplitude ratio $A_{\rm iso}/A_s$ and the spectral tilt $n_{\rm iso}$, for each instrument considered in this work.
The largest value we show for $A_{\rm iso}/A_s$ coincides with the upper bound from \textit{Planck}~\cite{Planck2018X}.
An Earth-based instrument tapping into the dark ages can improve on the \textit{Planck} constraints by about an order of magnitude, depending on the spectral index.
The smaller lunar array, LRAI, would significantly increase the significance of such constraints, while the larger LRAII could improve the constraints by two orders of magnitude for very low value of $n_{\rm iso}$ and several orders more for higher tilts.

The sensitivity of 21-cm dark ages measurements would be better than that for future galaxy surveys.
A large scale structure survey like Euclid allows, at best, to detect ($A_{\rm iso}/A_s=0.09, n_{\rm iso}=3$), while using an idealized version of MegaMapper could detect ($A_{\rm iso}/A_s=0.015, n_{\rm iso}=3$; $A_{\rm iso}/A_s=0.002, n_{\rm iso}=4$)~\cite{chung2023}.

\subsection{Running parameters}
\label{sec:runningresults}
We perform a Fisher forecast on the standard cosmological parameters~$\{ \omega_b,\omega_c,h,n_s,\ln 10^{10}A_s \}$, plus the running parameters~$\{\alpha_s, \beta_s\}$.
For fiducial values we take those of single-field slow-roll (SFSR) inflation: $\{\alpha_s,\beta_s \} = \{10^{-3},10^{-5} \}$~\cite{Kosowsky_1995, Cabass_2016}.
We then perform two analyses: one varying all the parameters above and one fixing cosmological ones to their \textit{Planck} 2018 best fit values~\cite{Planck2018VI}.

Table~\ref{tab:errors_running} shows predicted constraints for different instruments, indicating that the small lunar instrument, LRAI, could potentially test SFSR inflation through measurements of the $\alpha_s$ parameter, while the uncertainty on $\beta_s$ would be too large.
On the other hand, with the LRAII configuration, there could be a several sigma detection of (or a robust challenge to) $\alpha_s$ for the SFSR scenario.
LRAII would also have the sensitivity to start probing $\beta_s$ for SFSR.
As a point of comparison, the uncertainties from \textit{Planck} 2018 data are $\sigma \simeq 0.012$ for both parameters.

In the case where we vary cosmological parameters (first two rows in Table~\ref{tab:errors_running}), an SKAO-like experiment would be competitive with \textit{Planck} for $\alpha_s$ and improve by an order of magnitude for $\beta_s$.
Both lunar array cases would improve by 1 or more orders of magnitude over current CMB constraints, with LRAII being able to test the SFSR inflation scenario.

We also perform the Fisher analysis, fixing the standard $\Lambda$CDM parameter to demonstrate what the 21-cm experiments could achieve in principle.
Additionally, by the time a lunar array would be built, it is possible that very strong prior could be placed on the cosmological parameters from future precision CMB experiments.
In this case, an Earth-based instrument could start testing the SFSR scenario with $\alpha_s$.
Lunar arrays could measure $\alpha_s$ with extreme precision, while LRAII is needed to probe SFSR with $\beta_s$.
With the aid of future very precise priors, 21-cm IM surveys may be able to put inflationary models under strong scrutiny.

\begin{table}[t]
  \centering
  \begin{tabular}{|l|c|c|c|}
    \hline
    & SKAO-like & LRAI & LRAII \\
    \hline
    $\alpha_s, \rm{varying \,\,\, cosmology}$ & $1.3 \times 10^{-2}$ & $1.3 \times 10^{-3}$ & $1.1 \times 10^{-4}$ \\
    \hline
    $\beta_s, \rm{varying \,\,\, cosmology}$ & $3.7 \times 10^{-3}$ & $5.1 \times 10^{-4}$ & $2.1 \times 10^{-5}$ \\
    \hline
    $\alpha_s, \rm{fixed \,\,\, cosmology}$ & $1.3 \times 10^{-3}$ & $2.0 \times 10^{-4}$ & $3.5 \times 10^{-6}$ \\
    \hline
    $\beta_s, \rm{fixed \,\,\, cosmology}$ & $1.5 \times 10^{-3}$ & $2.0 \times 10^{-4}$ & $2.0 \times 10^{-6}$ \\
    \hline
  \end{tabular}
  \caption{Results from Fisher analyses for the first and second running parameters of the spectral index, $\alpha_s$ and $\beta_s$, respectively.
    We provide the $1\sigma$ errors with varying cosmological parameters (first two rows) and fixed cosmological parameters (bottom to rows), to be compared with \textit{Planck} 2018 ($\simeq 0.01$ for both parameters).
    The fiducial values for the running parameters are taken to be the standard SFSR $\{\alpha_s,\beta_s \} = \{10^{-3},10^{-5} \}$.}
  \label{tab:errors_running}
\end{table}

As a point of comparison, Ref.~\cite{Munoz2017} has forecasts for the spectral running parameters for future CMB and LSS experiments.
While an Earth-based interferometer would be competitive with future CMB measurements, it would not provide strong improvements.
The lunar arrays, however, would be more sensitive than future CMB+galaxy surveys, especially for the $\beta_s$ parameter, for which the improvement could be orders of magnitude.


\section{Conclusions}
\label{sec: conclusion}

The primordial power spectrum is well-constrained by CMB and LSS data to be almost scale invariant (and slightly red tilted) at large scales.
Several new-physics and dark matter models can imprint small-scale deviations from $\Lambda$CDM and are still largely unconstrained.
Many of these models induce an increase of power at smaller scales, producing a blue-tilted power spectrum.

In this work we investigate how future 21-cm intensity mapping interferometers targeting the dark ages can set limits on the enhancement of the power spectrum at small scales.
We consider three example instruments: a future Earth-based experiment that could detect the end of the dark ages (SKAO-like), and two radio arrays on the far side of the Moon (LRAI, LRAII).
LRAI is similar to an SKAO extension but on the Moon; LRAII covers almost the entirety of the Moon's far side with radio antennas, as proof of principle for what can be achieved.
For each instrument, we determine the signal-to-noise ratio for detecting a blue-tilted primordial power spectrum that arises from an enhancement of small-scale curvature modes and cold dark matter isocurvature modes.

For enhanced curvature modes, the most important instrumental feature is the baseline of the interferometer, which translates into the maximum multipole that the instrument can probe.
We find that any blueness of the spectrum is detectable, as long as the growth starts within the probed multipole range.
Thus, while the significance of the detection would be higher for a small lunar array, there would be not much difference from an Earth-based instrument with the same baseline.
We analyze instruments with baselines of $100 \, {\rm km}$ and and $3000 \, {\rm km}$, which translates into detecting any blue-tilted deviation that occurs at scales larger than $k\sim 8 \, {\rm Mpc}^{-1}$ and $k\sim 230 \, {\rm Mpc}^{-1}$, respectively.

For small-scale enhancements from cold dark matter isocurvature modes, even just an Earth-based instrument would have greater sensitivity to a blue tilt than \textit{Planck} or future galaxy surveys.
Lunar arrays would further increase the significance of detection and, in the case of the LRAII, improve upon \textit{Planck} constraints by several orders of magnitude.

Finally, we perform a Fisher analysis to forecast constraints on the running of the spectral tilt.
Our results show that, while a Earth-based instrument can be competitive with current and near-future CMB experiments, lunar arrays could improve such constraints by one or more orders of magnitude, for both running parameters, depending on the priors we choose for the standard $\Lambda$CDM parameters.
The larger LRAII could, even with no external priors, not only set very stringent constraints on the running parameters, but also probe the expected values for standard single-field slow-roll inflation.

In conclusion, interferometers capable of measuring the 21-cm signal from the cosmic dark ages would be able to detect blue spectra that could originate from several extensions to the standard cosmological and particle physics models.
While we have focused on the scale and slope of the power spectrum increase, we note that these parameters can be mapped to specific inflationary and particle models.
Given the range of scales that can we probed by such instruments, they could provide invaluable and otherwise unobtainable information on exotic physics and test inflationary models to unprecedented precision.


\begin{acknowledgments}
A.R. thanks the University of Texas, Austin, for the hospitality during the initial stages of this work.
K.B. thanks the University of Padova for the hospitality during the intermediate stages of this work.
We thank Mustafa Amin, Sabino Matarrese, Nicola Bellomo, Andrew Long, and Julian Mu\~{n}oz for useful discussions.

K.B. acknowledges support from the National Science Foundation (NSF) under Grant No.~PHY-2112884 and acknowledges the ``Dark Matter Theory, Simulation, and Analysis in the Era of Large Surveys'' workshop and the Kavli Institute for Theoretical Physics (KITP) for the hospitality and support under NSF Grant No.~PHY-2309135.
A.R. acknowledges funding from the Italian Ministry of University and Research
(MIUR) through the “Dipartimenti di eccellenza” project “Science of the Universe”.
\end{acknowledgments}

\appendix

\section{SNR dependence on instrument specifications}
\label{sec:Appendix}

\begin{figure}
  \includegraphics[width=.49\textwidth]{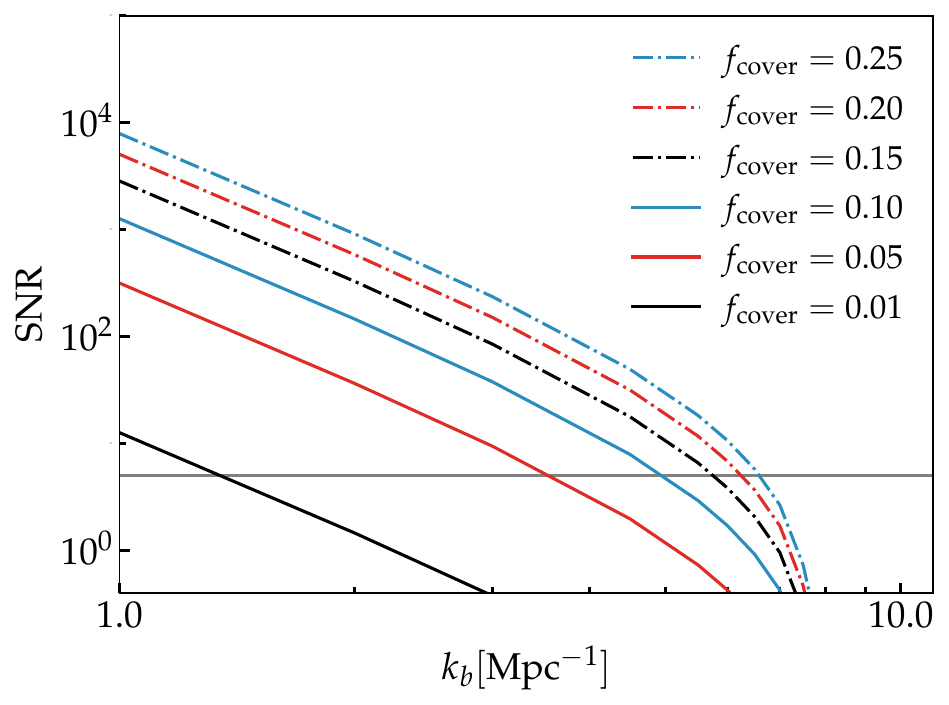}
  \includegraphics[width=.49\textwidth]{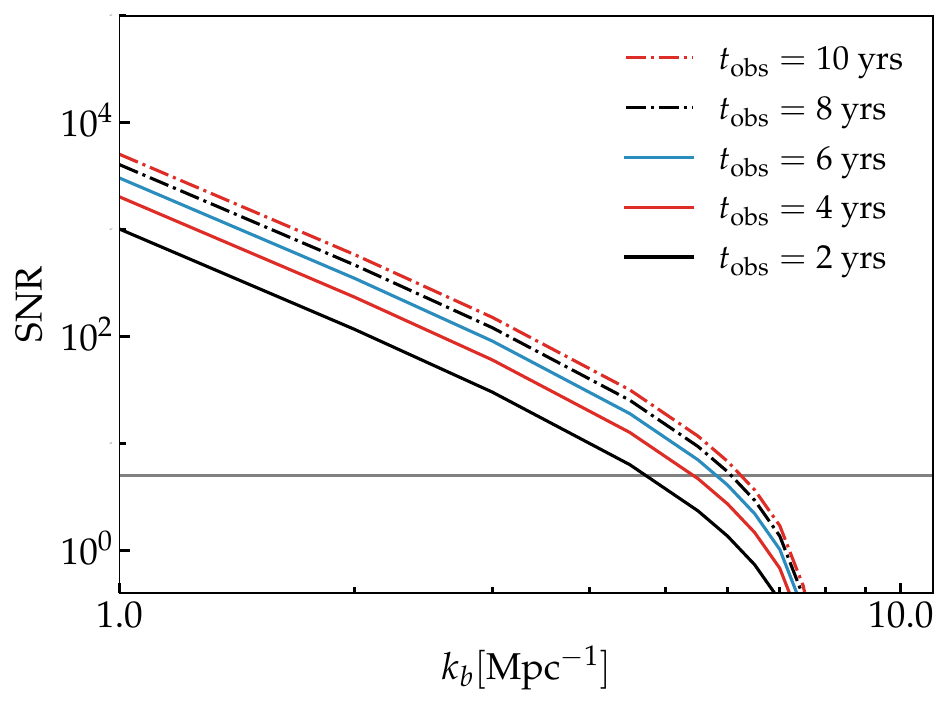}
  \caption{SNR as a function of $k_{b}$ for a blue-tilted deviation from $\Lambda$CDM, using the SKAO-like instrument as specified in Table~\ref{tab:survey_specs} but with different values for $f_{\rm cover}$ (top) and $t_{\rm obs}$ (bottom), as indicated in the captions. We set $n_b =3$ and indicate $\rm{SNR}=5$ with a gray horizontal line for reference.}
  \label{fig:SNR different SKAO}
\end{figure}

\begin{figure}
  \includegraphics[width=.49\textwidth]{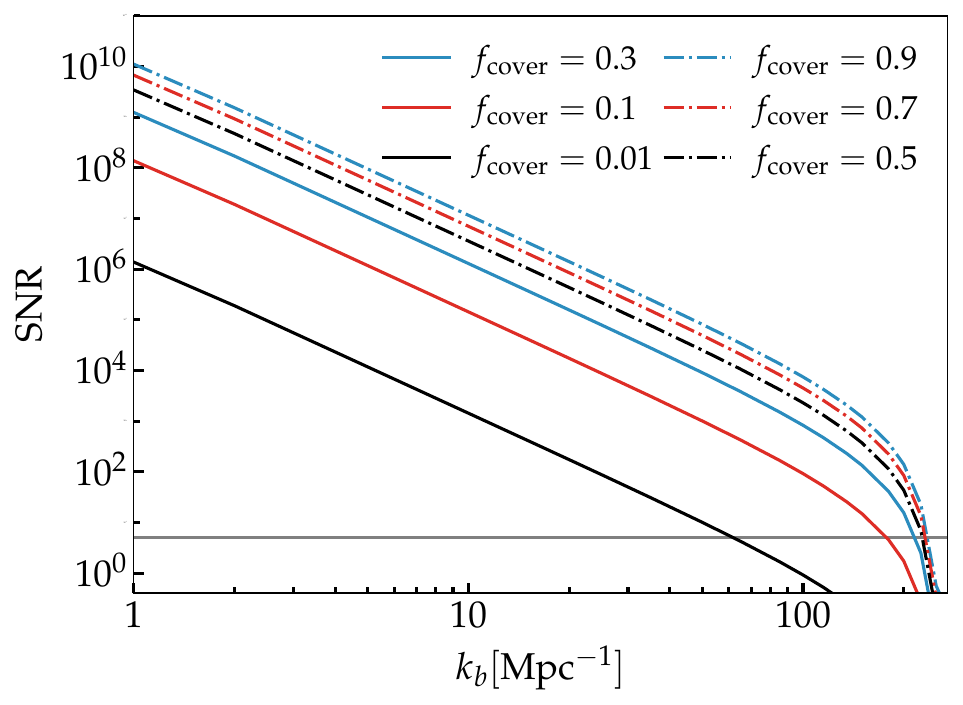}
  \includegraphics[width=.49\textwidth]{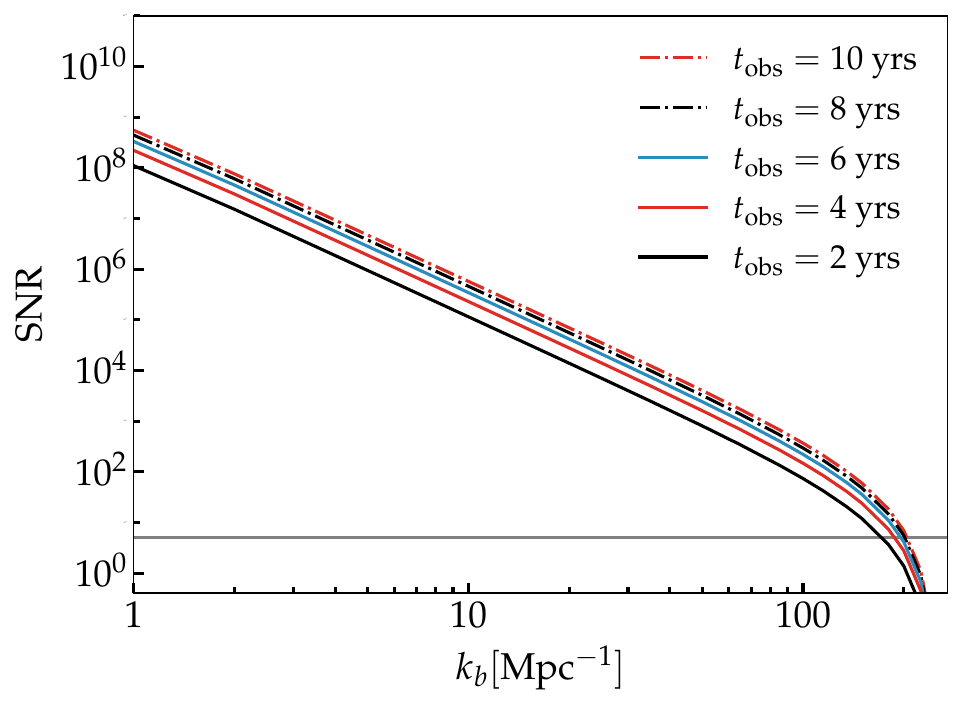}
  \caption{SNR as a function of $k_{b}$ for a blue-tilted deviation from $\Lambda$CDM, using the LRAII instrument as specified in Table~\ref{tab:survey_specs} but with different values for $f_{\rm cover}$ (top) and $t_{\rm obs}$ (bottom), as indicated in the captions. We set $n_b =3$ and indicate $\rm{SNR}=5$ with a gray horizontal line for reference.}
  \label{fig:SNR different LRAII}
\end{figure}

In this Appendix, we consider how the total observational time $t_{\rm obs}$ and the fraction of the array area covered by antennas $f_{\rm cover}$ affects the SNR for SKAO-like (Fig.~\ref{fig:SNR different SKAO}) and LRAII (Fig.~\ref{fig:SNR different LRAII}). 
For both instruments, it is clear that even with a much shorter running time, the obtained signal is still very strong. Figure~\ref{fig:SNR different SKAO} shows that for SKAO-like, we need at least $f_{\rm cover}=0.2$ in order not to lose sensitivity on the smallest scales. Similarly for LRAII in Fig.~\ref{fig:SNR different SKAO}, we require $f_{\rm cover} \gtrsim 0.3$ to probe the smallest scales.

\bibliography{bibliography}

\end{document}